\documentstyle[12pt]{article}
\renewcommand{\thefootnote}{\fnsymbol{footnote}}
\newlength{\extraspace}
\newlength{\extraspaces}
\newcommand{\be}{\begin{equation}
\addtolength{\abovedisplayskip}{\extraspaces}
\addtolength{\belowdisplayskip}{\extraspaces}
\addtolength{\abovedisplayshortskip}{\extraspace}
\addtolength{\belowdisplayshortskip}{\extraspace}}
\newcommand{\ee}{\end{equation}}
\newcommand{\bea}{\begin{eqnarray}
\addtolength{\abovedisplayskip}{\extraspaces}
\addtolength{\belowdisplayskip}{\extraspaces}
\addtolength{\abovedisplayshortskip}{\extraspace}
\addtolength{\belowdisplayshortskip}{\extraspace}}
\newcommand{\eea}{\end{eqnarray}}
\def\Journal#1#2#3#4{{#1} {\bf #2} (#3) #4}

\def\NPB{{\em Nucl. Phys.} B}

\def\PLB{{\em Phys. Lett.}  B}
\def\PRL{\em Phys. Rev. Lett.}

\def\AAM{\em Adv. Appl. Math.}
\def\CMP{\em Commun. Math. Phys.}
\def\JMP{\em J. Math. Phys.}
\def\MPLA{{\em Mod. Phys. Lett.} A}

\newcommand{\nonu}{\nonumber}
\newcommand{\tr}{\, {\rm tr} \,}
\newcommand{\e}{\, {\rm e}}
\newcommand{\ie}{{\it i.e.}\ }

\begin{document}
\addtolength{\baselineskip}{.5mm}
\thispagestyle{empty}
%
\begin{flushright}
OU--HET 240 \\
March, 1996 \\
\end{flushright}
\vspace{10mm}
\begin{center}
{\Large{\bf{ Exact Solution of 1-matrix Model
\footnote{Based on a contribution paper for the proceedings of
the workshop "Frontiers in Quantum Field Theory" celebrating
Professor Kikkawa's 60th birthday, held at Osaka
University, Osaka, Japan, December 14-17, 1995.}
}}}
\\[29mm]
{\sc Hiroshi Shirokura}\footnote{A JSPS Research Fellow,\quad
E-mail address: siro@funpth.phys.sci.osaka-u.ac.jp} \\[8mm]
{\it Department of Physics, Osaka University \\[3mm]
Toyonaka, Osaka 560, JAPAN} \\[44mm]
%
{\bf ABSTRACT}\\[9mm]
{\parbox{13cm}{\hspace{5mm}
I review my new method for solving general 1-matrix models by expanding
in $N^{-1}$ without taking a physical continuum limit.
Using my method, each coefficient of the free energy in the genus
expansion is exactly computable.
One can include physical information in a function which is uniquely
specified by the action of the model.
My method gives completely the same result with the usual one
if the physical fine tuning is done and the leading singular terms are
extracted.
I also calculate in the genus three case and confirm the validity of
my method.
}}
\end{center}
\vfill
\newpage
\renewcommand{\thefootnote}{\arabic{footnote}}
\setcounter{equation}{0}
\setcounter{footnote}{0}
%

In this review article, I want to explain a new method for solving
1-matrix models that I have found recently \cite{SIRO}.
Matrix models have been vigorously studied as an exactly solvable
toy model for two dimensional gravity.
Even though 1-matrix models are the easiest of all matrix models,
only the planar limit \cite{BIPZ} have been argued mainly
until the appearance of the double scaling limit \cite{GRMI,BRKA}.
This limit enables us to convert a recursive equation derived by
the orthogonal polynomial method and large-$N$ expansion
($N$ is the size of matrix)
to a non-linear differential equation for the specific heat.
The solution of this differential equation contains non-perturbative
effects which are absent in the perturbative expansion by the Feynman
diagrams.
My method gives another way to solve the recursive equation.
Instead of taking the double scaling limit, I expand the recursive
equation in $N^{-1}$ and solve it order by order.
What an intuitive way!
However, the equations for the expansion coefficients of $r_\epsilon$
($r_\epsilon$ will be explained shortly later) derived in this way
is hard to solve even when the expansion order is low.
The obstacle to advance was the fact that these equations have
summations over paths \cite{BIZ}.
To overcome this difficulty, I found a beautiful relation between
these summations and a function $W(r)$ which is specified uniquely
by the action of the model.
Using this relation I find the expansion coefficients of free energy
until genus three \cite{SIRO}.
In appendix I will show the explicit result for the genus three case.
In principle one can continue this calculation to higher genus.
However, the calculation will become rapidly complicated for large
genus.

%
I follow the notations of my previous paper \cite{SIRO}.
Since in this short review I cannot explain all of the notations,
I will sometimes omit explanation for some undefined signs.
Consult \cite{SIRO} for further information.
A 1-matrix model is defined by an integral over an $N\times N$
hermitian matrix $\Phi$,
\bea
  Z_N(\{\lambda\}) & =  & \int\!\!{\cal D}\Phi \e^{-NS[\Phi,\{\lambda\}]}
                          \,, \nonu \\
  S[\Phi, \{\lambda\}] & = &
   \tr \left[
         \frac{1}{2}\Phi^2+\sum_{p=1}^\infty\lambda_p\Phi^{2p+2}
       \right]\,,
\label{eqn:1MatrixModel}
\eea
where $\{\lambda\}$ denotes the set of coupling parameters.
The main purpose of the game is to find the expansion coefficients of
the free energy
\be
  E_N(\{\lambda\}) = \frac{1}{N^2}\log\frac{Z_N(\{\lambda\})}{Z_N(0)}
                   = \sum_{h=0}^\infty N^{-2h}\e^{(h)}(a^2, \{\lambda\})\,.
\label{eqn:GenusExpansion}
\ee
In (\ref{eqn:GenusExpansion}) I indicated $a^2$-dependence of $e^{(h)}$.
This parameter $a^2$ is a real solution of $W(r)=1$
\footnote{If such a solution does not exist, the matrix integration
will be complex and loses its physical meaning.}, where $W(r)$ is
the function I mentioned before
\be
  W(r) = r\left[
             1+\sum_{p=1}^\infty\frac{(2p+2)!}{p!(p+1)!}\lambda_pr^p
          \right]\,.
\label{eqn:CharacteristicFunction}
\ee

%
Using the orthogonal polynomial and the Euler-Maclaurin formula
the contribution from genus $h$ is described as follows
\bea
  \e^{(h)} & = & \int_0^1\!\!dx(1-x)\tilde{r}_{2h}(x)+K_{2h} \nonu \\
           &   & -\sum_{s=1}^h(-1)^s\frac{B_s}{(2s)!}
                 \left.
                  \left\{(1-x)\tilde{r}_{2(h-s)}(x)\right\}^{(2s-1)}
                 \right|_0^1\,,
\label{eqn:GenusH}
\eea
where $B_s$ are the Bernoulli numbers.
This formula is my starting point.
The universal contribution which is survived in the continuum limit
lies in the first term on the right hand side.
In (\ref{eqn:GenusH}) $\tilde{r}_{2s}$ are expansion coefficients
of the function
\be
\tilde{r}_\epsilon(x) \equiv \log\frac{r_\epsilon(x)}{x}
= \sum_{s=0}^\infty\epsilon^{2s}\tilde{r}_{2s}(x)\,,
\ee
where $\epsilon=N^{-1}$.
Since the boundary term $K_{2h}$ can be derived from the function $W(r)$,
what we need to compute $e^{(h)}$ is to solve a recursive equation
for $r_\epsilon(x)$ \cite{BIZ}
\be
  x = r_\epsilon(x)
      \left[
       1+\sum_{p=1}^\infty2(p+1)\lambda_p\sum_{l\in L_p}\prod_{i=1}^p
       r_\epsilon(x+s_i(l)\epsilon)
      \right]\,,
\label{eqn:RecursiveEq}
\ee
from which one can find the moments $\tilde{r}_{2s}$ in (\ref{eqn:GenusH}).
In (\ref{eqn:RecursiveEq}) $L_p$ is a set of all paths with length
$(2p+1)$ discussed in my previous paper and $-p\leq s_i(l)\leq p$
are $p$ integers assigned to each path $l$.
These so-called "height" variables are connected to the set of all
combinations which pick up $p$ elements out of a set of $(2p+1)$
integers $\{1,2,\ldots,2p+1\}$.
If one of the combination is $\{\sigma_1,\ldots,\sigma_p\}$\,,
$(\sigma_1\leq\cdots\leq\sigma_p)$,
each integer $\sigma_i$ is related to a height variable for a
path $l \in L_p$ according to the following simple rule:
\be
s_i=\sigma_i-2i,\quad i=1,2,\ldots,p\,.
\ee
If the action of the model is suitably fine tuned and the double scaling
limit is taken, this recursive equation turns to the string equation.
Alternatively I solve (\ref{eqn:RecursiveEq}) by expanding as
$r_\epsilon(x) = \sum_s \epsilon^{2s}r_{2s}(x)$.
One should note that the equation for $r_0$ is $x=W(r_0)$.
Thus the boundary $x=1$ in (\ref{eqn:GenusH}) corresponds to
$r_0=a^2$.
This boundary contributes to the non-analyticity of $e^{(h)}$
and the another boundary $x=0$ to the regular part in terms of
coupling parameters.

%
It is convenient to introduce a number $I_p(n_1,\ldots,n_m)$
specified by a Young's diagram $(n_1,\ldots,n_m)$ whose
number of rows is $m$.
This Young's diagram corresponds to a partition of a positive even
integer $2n$, \ie, $\sum n_i=2n$.
These numbers play an important role in my method,
for they appear in the expanded recursive equations derived from
(\ref{eqn:RecursiveEq}).
If $f(x)$ is an arbitrary infinitely differentiable function,
the number is defined by a following Taylor expansion
\[
  \sum_{l\in L_p}\prod_{i=1}^pf(x+s_i(l)\epsilon)
  = \sum_{n=0}^\infty \epsilon^{2n}F_{2n}(p)\,,
\]
\be
  F_{2n}(p) = \sum_{(n_1,\ldots,n_m)\in Y_{2n}} I_p(n_1,\ldots,n_m)
              f^{(n_1)}(x)\cdots f^{(n_m)}(x)\{f(x)\}^{p-m}\,,
\label{eqn:DefOfSum}
\ee
where $Y_{2n}$ is the set of partitions of $2n$.
The number vanishes when $p<m$.
These numbers depend on $p$ which parameterizes the path length.
I have decided the $p$-dependence of these numbers guided by
two kinds of arithmetic generating functions:
\bea
{}_{2p+1}C_p(x)
& = &
\sum_{n=0}^\infty\sum_{(n_1,\ldots,n_m)\in Y_{2n}}I_p(n_1,\ldots,n_m)
x^{2n}\,, \\
\Gamma_p(x)
& = &
\sum_{n=0}^\infty I_p(2n)x^{2n}\,.
\eea
Remarkably these generating functions are both described by
$q$-deformed integers, where $q=\exp(x/2)$.
The answer for this problem can be presented in a simple way.
To do this it is convenient to define a generating function for
these numbers
\be
  I(r_0;(n_1,\ldots,n_m)\,) = \sum_{p=m}^\infty 2(p+1)\lambda_p
                              I_p(n_1,\ldots,n_m){r_0}^{p-m}\,.
\label{eqn:GenFuncForIP}
\ee
The $p$-dependence of $I_p(n_1,\ldots,n_m)$ is described by
the following remarkable relation between the generating function and
the function $W(r_0)$,
\be
  I(r_0;(n_1,\ldots,n_m)\,) = \sum_{i=0}^{n-1}\alpha_i(n_1,\ldots,n_m)
                              r_0^iW^{(i+m+1)}(r_0)\,,
\label{eqn:RelIandW}
\ee
where $\alpha_i(n_1,\ldots,n_m)$ are positive rational numbers.
These $n$ rational numbers are completely fixed by $n$ explicit
values of $I_p(n_1,\ldots,n_m)$ for small $p$.
Using a computer these values are computable.
One meets the generating functions defined in (\ref{eqn:GenFuncForIP})
and their derivatives with respective to $r_0$
when the large-$N$ expansion of (\ref{eqn:RecursiveEq}) is performed.
For example, one obtains a equation for $r_2$ (rather for
$\tilde{r}_2=r_2/r_0$) from the $\epsilon^2$ order,
\be
 r_2W'(r_0)+r_0
 \left\{I(r_0;(2)\,){r_0}^{(2)}+I(r_0;(1,1)\,)({r_0}^{(1)})^2\right\}
 = 0\,.
\ee
The relation (\ref{eqn:RelIandW}) is easily verified for the Young's
diagrams $(2)$, and $(1,1)$:
\[
 I(r_0;(2)\,)=\frac{1}{6}W^{(2)}(r_0)\,,\quad
 I(r_0;(1,1)\,)=\frac{1}{12}W^{(3)}(r_0)\,.
\]
Since one can write any derivatives $r_0^{(n)}$ in terms of
derivatives of $W(r_0)$ using the relation $x=W(r_0)$,
the moment $\tilde{r}_2$ can be described only in terms of $W(r_0)$.

These expanded recursive equations give the solutions for
the expansion coefficients of $\tilde{r}_\epsilon$.
These coefficients are always written in the following form for
$h\geq 1$,
\be
\tilde{r}_{2h}=
\left(\frac{1}{W'(r_0)}\frac{d}{dr_0}\right)^2
\sum_{s=0}^{h-1}(-1)^sr_0^s\Omega_s^{(h)}(r_0)\,.
\label{eqn:CoeffTildR}
\ee
The above form has an important meaning:
the integral in (\ref{eqn:GenusH}) can be explicitly performed and
only the boundaries $x=0, 1$ contribute to the free energy.
I must introduce a subset $X_k \subset Y_{2k}$,
where $Y_{2k}$ is the set of partitions of $2k$ to explain
the functions $\Omega_s^{(h)}$ in (\ref{eqn:CoeffTildR}).
This subset is defined as a set of the partitions of $2k$ whose number
of rows of the corresponding Young's diagram are $k$,
\be
X_k = \{ (n_1,\ldots,n_k) | \sum_in_i=2k \}\,.
\ee
Then I define sets of rational polynomial functions.
If $f(x)$ is an infinitely differentiable function, the sets are
defined as follows
\be
\mbox{\boldmath$Q$}_{k,s}(f(x)) \equiv
\bigoplus_{(n_1,\ldots,n_k) \in X_k} \mbox{\boldmath$Q$}
\left\{\prod_{i=1}^kf^{(n_i)}\right\}/(f')^s\,.
\ee
Then the functions $\Omega_s^{(h)}(r_0)$ are elements of such sets,
\be
\Omega_s^{(h)}(r_0) \in
\mbox{\boldmath$Q$}_{2(h-1)+s,4(h-1)+s}(W(r_0))\,.
\label{eqn:Omega}
\ee

%
The actual computation process of $e^{(h)}$ is quite boring,
so I show only the result.
Since the spherical contribution $e^{(0)}$ is well-known,
I omit this case.
The genus one contribution is surprisingly simple and is
exceptional comparing to higher genus cases,
\be
e^{(1)} = -\frac{1}{12}\log \{a^2W'(a^2)\}\,.
\label{eqn:E1}
\ee
For $h\geq 2$ the explicit form of $e^{(h)}$ is very complicated,
but they are always written in a following form,
\be
  \e^{(h)} = \sum_{s=-2(h-1)}^{h-1} E_s^{(h)}(a^2)a^{2s}
             +(-1)^h\frac{B_h}{4h(h-1)}\,,
\label{eqn:GenBehavForEH}
\ee
where the coefficients $E_s^{(h)}(r_0)$ are the same kind of rational
polynomial functions of derivatives of $W(r_0)$ as $\Omega_s^{(h)}$
except that the subscript $s$ might be negative,
\be
E_s^{(h)}(r_0) \in
\mbox{\boldmath$Q$}_{2(h-1)+s, 4(h-1)+s}(W(r_0))\,.
\label{eqn:RuleEsh}
\ee
Here the coefficient with the greatest $s$ is equivalent to
$\Omega_{h-1}^{(h)}$.
The Contributions from negative $s$ are understood through a relation,
\be
\sum_{s=-2(h-1)}^{-1}E_s^{(h)}(a^2)a^{2s} =
(-1)^h\frac{(2h-1)B_h}{(2h)!}
\left(\frac{1}{W'(a^2)}\frac{d}{da^2}\right)^{2h-2}\log a^2\,.
\ee
Furthermore $E_0^{(h)}$ can be constructed from the following relation,
\be
E_0^{(h)}(0) = -\lim_{r\rightarrow 0}
\left[
 \sum_{s=-2(h-1)}^{-1}E_s^{(h)}(r)r^s+
 (-1)^h\frac{B_h}{4h(h-1)}\frac{1}{W(r)^{2h-2}}
\right]\,.
\ee
The most important thing to note in (\ref{eqn:GenBehavForEH}) is
that for $h\geq 1$, the $x=0$ boundary contributions are offset
and there is no term that depends on the derivatives of
$W(r_0)$ at $r_0=0$.
The origin of this cancellation of boundary contributions is still
mysterious.
In appendix I concretely compute $e^{(3)}$.
The reader will understand how complicated the actual form of
the free energy is.

%
One must fine tune the action to treat the 1-matrix model defined by
(\ref{eqn:1MatrixModel}) as a model of two dimensional gravity.
The critical exponent and the physical observables are included in
this model by the fine tuning.
I think that the most heuristic way for the fine tuning
is simply described as follows
\be
  S_{k,b}[\Phi, \{\tau\}] = N + \sum_{i=-2}^m\tau_i
                                \rho_{n_i,k}^{M_i,b}[\Phi]\,.
\label{eqn:PhysFineTuning}
\ee
In (\ref{eqn:PhysFineTuning}) $\rho_{n_i,k}^{M_i,b}[\Phi]$ expresses
a discretized scaling operator.
The $i=0$ term corresponds to the cosmological term,
$(n_0,M_0,\tau_0) \equiv (0,0,t)$.
The terms with negative $i$ specify the criticality of the system,
$(n_{-1},M_{-1},\tau_{-1})=(k,-1,b)$,
$(n_{-2},M_{-2},\tau_{-2})=(k,-2,-1)$.
The integer $k$ is related to the string susceptibility
$\gamma = -1/k$ in the continuum theory.
The terms with positive $i$ describe source terms for scaling operators
and $n_i/k$ is equivalent to the gravitational dimension of the $i$-th
scaling operator.
The superscripts $M_i$ and $b$ are non-universal counterparts of $n_i$
and $k$ respectively.
Since the formal gravitational dimension for the $(-1)$-th scaling
operator is $1$, this term corresponds to a marginal operator
and $b$ can be interpreted as a marginal deformation parameter.
Though I do not show the explicit form of the discretized version of
the scaling operators, it is sufficient to show the converted function
$ W_{k,b}(r,\{\tau\}) $ derived from the fine tuned action
(\ref{eqn:PhysFineTuning}),
\be
  W_{k,b}(r, \{\tau\}) = 1-\sum_{i=-2}^m\left(\frac{r}{k+b}\right)^{M_i+2}
                         \left(1-\frac{r}{k+b}\right)^{n_i}\tau_i\,.
\label{eqn:PhysW}
\ee
Using this function, one can control even the non-universal information
systematically.
For example, one can find that the regular part of the spherical free
energy is at most a bilinear function with respective to the couplings
$\tau_i, (i=0,1,\ldots)$.

%
Now I fine tune the action to handle the free energy
(\ref{eqn:GenusExpansion}) as a generating function for
correlation functions of physical observables.
From the physical point of view a continuum limit must be taken.
The rule for extracting the universal terms from the result
of $e^{(h)}$ (\ref{eqn:GenBehavForEH}) is quite simple:
replace the derivatives of $W(r_0)$ at $r_0=a^2$
in the term $E_{h-1}^{(h)}(a^2)a^{2(h-1)}$ according to the following
manner,
\be
  a^{2s}W^{(s)}(a^2) \longrightarrow
  \left( -\frac{1}{f_0^{(1)}}\frac{d}{dt} \right)^{s-1}
  \frac{1}{f_0^{(1)}}\,,
\label{eqn:ExtractRule}
\ee
where $f_0(t)$ is the solution of
$ t = f_0^k-\sum_i\tau_if_0^{n_i} $.
Note that at this stage the non-universal parameters $M_i$ and $b$
disappear.
When all sources are turned off, this function $f_0(t)$ corresponds to
the specific heat in the continuum limit.
Finally the universal contribution $e_0^{(h)}(t)$ becomes a rational
function of the derivatives of $f_0(t)$,
\be
e_0^{(h)}(t) \in \mbox{\boldmath$Q$}_{3(h-1), 4(h-1)}(f_0(t))\,,
\quad (h\geq 2)\,.
\ee
One can compute any correlation functions of physical observables
from this universal function $e_0^{(h)}(t)$.
Their $t$-dependence completely agrees with the prediction of the
continuum theory.
When all source terms are absent, one can confirm that the
$t$-dependence of the free energy actually coincides with
the result of the Liouville theory \cite{DDK},
\be
\left.
e_0^{(h)}(t)
\right|_{\{\tau\}=0}
\propto t^{2-\gamma(h)}\,,
\ee
where,
\bea
2-\gamma(h) & = &\frac{1}{k} \cdot 3(h-1) - 6(h-1)
                -4(h-1)\left(\frac{1}{k}-1\right) \nonu \\
            & = & (1-h)(2-\gamma)\,.
\eea
In this way my method reproduces all of the known results.

At the end of this short article, one should note that the matrix
size $N$ is still finite.
Discarding except for the most singular terms corresponds to taking
the double scaling limit,
$N\rightarrow\infty,\ t\rightarrow 0$ and
$N^2t^{2-\gamma}\rightarrow\mbox{const.}$

%
\section*{Appendix}

In this appendix I give detailed results for genus three.
The starting point for  genus three is read from (\ref{eqn:GenusH}),
\bea
e^{(3)} & = & \int_0^1\!\!dx(1-x)\tilde{r}_6 + K_6 \nonu \\
        &   & +\frac{1}{12}
               \left.\left((1-x)\tilde{r}_4\right)^{(1)}\right|_0^1
              -\frac{1}{720}
               \left.\left((1-x)\tilde{r}_2\right)^{(3)}\right|_0^1
               \nonu \\
        &   & +\frac{1}{30240}
               \left.\left((1-x)\tilde{r}_0\right)^{(5)}\right|_0^1\,,
\label{eqn:E3}
\eea
where $K_6$ is given by derivatives of $W(r_0)$ at $r_0=0$,
\bea
K_6
& = &
-\frac{1}{5760}
 [3714{W^{(2)}(0)}^5-6340{W^{(2)}(0)}^3W^{(3)}(0)
 +1823W^{(2)}(0){W^{(3)}(0)}^2 \nonu \\
&   &
 +1455{W^{(2)}(0)}^2W^{(4)}(0)-318W^{(3)}(0)W^{(4)}(0)
 -210W^{(2)}(0)W^{(5)}(0) \nonu \\
&   &
 +15W^{(6)}(0)]\,.
\eea
The solutions for $\tilde{r}_2=r_2/r_0$ and
$\tilde{r}_4=r_4/r_0-(r_2/r_0)^2/2$ were derived in my previous work.
Thus the only unknown function in (\ref{eqn:E3}) is
\be
\tilde{r}_6=\frac{r_6}{r_0}-\frac{r_2}{r_0}\frac{r_4}{r_0}
+\frac{1}{3}\left(\frac{r_2}{r_0}\right)^3\,.
\ee

To find this function according to my procedure we need the
generating functions for the partitions of until six defined in
(\ref{eqn:GenFuncForIP}).
One should rely on a computer to do this.
Since the generating functions for the partitions of two and four
have been implicitly derived in my previous work,
I show here only the functions for eleven partitions of six
\footnote{Here $I(n_1,\ldots,n_m)$ is an abbreviation
for $I(r_0;(n_1,\ldots,n_m))$.}:
\bea
I(6)
& = &
 \frac{1}{2160}W^{(2)}+\frac{1}{360}r_0W^{(3)}
+\frac{1}{840}r_0^2W^{(4)}\,, \\
I(5,1)
& = &
 \frac{7}{720}W^{(3)}+\frac{13}{840}r_0W^{(4)}
+\frac{1}{280}r_0^2W^{(5)}\,, \\
I(4,2)
& = &
 \frac{23}{1440}W^{(3)}+\frac{17}{560}r_0W^{(4)}
+\frac{19}{2520}r_0^2W^{(5)}\,, \\
I(3,3)
& = &
 \frac{17}{2160}W^{(3)}+\frac{11}{630}r_0W^{(4)}
+\frac{23}{5040}r_0^2W^{(5)}\,, \\
I(4,1,1)
& = &
 \frac{31}{1008}W^{(4)}+\frac{181}{6048}r_0W^{(5)}
+\frac{5}{1008}r_0^2W^{(6)}\,, \\
I(3,2,1)
& = &
 \frac{113}{1260}W^{(4)}+\frac{737}{7560}r_0W^{(5)}
+\frac{43}{2520}r_0^2W^{(6)}\,, \\
I(2,2,2)
& = &
 \frac{11}{560}W^{(4)}+\frac{227}{10080}r_0W^{(5)}
+\frac{61}{15120}r_0^2W^{(6)}\,, \\
I(3,1,1,1)
& = &
 \frac{1}{24}W^{(5)}+\frac{13}{432}r_0W^{(6)}
+\frac{1}{252}r_0^2W^{(7)}\,, \\
I(2,2,1,1)
& = &
 \frac{23}{288}W^{(5)}+\frac{35}{576}r_0W^{(6)}
+\frac{83}{10080}r_0^2W^{(7)}\,, \\
I(2,1,1,1,1)
& = &
 \frac{1}{32}W^{(6)}+\frac{181}{10080}r_0W^{(7)}
+\frac{17}{8640}r_0^2W^{(8)}\,, \\
I(1,1,1,1,1,1)
& = &
 \frac{1}{448}W^{(7)}+\frac{1}{960}r_0W^{(8)}
+\frac{1}{10368}r_0^2W^{(9)}\,.
\eea
Then it is possible to solve the recursive equation for $\tilde{r}_6$
and the results are indeed written in the form of (\ref{eqn:CoeffTildR}),
\be
\tilde{r}_6 =
\left(
 \frac{1}{W'(r)}\frac{d}{dr_0}
\right)^2
[\Omega_0^{(3)}-r_0\Omega_1^{(3)}+r_0^2\Omega_2^{(3)}]\,.
\ee
Explicit form of the coefficients $\Omega_s^{(3)}$ obeys the rule
(\ref{eqn:Omega}),
\bea
\Omega_0^{(3)}
& = &
[18039(W^{(2)})^4
-24312W'(W^{(2)})^2W^{(3)} \nonu \\
&   &
+3182(W')^2(W^{(3)})^2
+5088(W')^2W^{(2)}W^{(4)} \nonu \\
&   &
-528(W')^3W^{(5)}]/(725760(W')^8) \,, \\
\Omega_1^{(3)}
& = &
[21000(W^{(2)})^5
-38535W'(W^{(2)})^3W^{(3)} \nonu \\
&   &
+12114(W')^2W^{(2)}(W^{(3)})^2
+9495(W')^2(W^{(2)})^2W^{(4)} \nonu \\
&   &
-2315(W')^3W^{(3)}W^{(4)}
-1494(W')^3W^{(2)}W^{(5)} \nonu \\
&   &
+119(W')^4W^{(6)}]/(181440(W')^9) \,, \\
\Omega_2^{(3)}
& = &
[34300(W^{(2)})^6
-81060W'(W^{(2)})^4W^{(3)} \nonu \\
&   &
+43050(W')^2(W^{(2)})^2(W^{(3)})^2
-2915(W')^3(W^{(3)})^3 \nonu \\
&   &
+22260(W')^2(W^{(2)})^3W^{(4)}
-13452(W')^3W^{(2)}W^{(3)}W^{(4)} \nonu \\
&   &
+607(W')^4(W^{(4)})^2
-4284(W')^3(W^{(2)})^2W^{(5)} \nonu \\
&   &
+1006(W')^4W^{(3)}W^{(5)}
+539(W')^4W^{(2)}W^{(6)} \nonu \\
&   &
-35(W')^5W^{(7)}]/(362880(W')^{10})\,.
\eea

As the function $\tilde{r}_6$ is explicitly obtained,
it is easy to compute the generating function (\ref{eqn:E3}).
It is certainly written in the form of (\ref{eqn:GenBehavForEH}),
\be
e^{(3)} = \sum_{s=-4}^2a^{2s}E_s^{(3)}(a^2) - \frac{1}{1008}\,.
\label{eqn:AnsE3}
\ee
The coefficients of (\ref{eqn:AnsE3}) are very complicated
but they indeed satisfy (\ref{eqn:RuleEsh}),
\bea
E_2^{(3)}
&\!\! = \!\!&
\Omega_2^{(3)}(a^2)\,, \\
E_1^{(3)}
&\!\! = \!\!&
-[21420{W^{(2)}(a^2)}^5
-40110W'(a^2){W^{(2)}(a^2)}^3W^{(3)}(a^2) \nonu \\
&\!\!   \!\!&
+12783{W'(a^2)}^2W^{(2)}(a^2){W^{(3)}(a^2)}^2
+10170{W'(a^2)}^2{W^{(2)}(a^2)}^2W^{(4)}(a^2) \nonu \\
&\!\!   \!\!&
-2488{W'(a^2)}^3W^{(3)}(a^2)W^{(4)}(a^2)
-1644{W'(a^2)}^3W^{(2)}(a^2)W^{(5)}(a^2) \nonu \\
&\!\!   \!\!&
+133{W'(a^2)}^4W^{(6)}(a^2)]/(362880{W'(a^2)}^9)\,, \\
E_0^{(3)}
&\!\! = \!\!&
[1575{W^{(2)}(a^2)}^4
-1800W'(a^2){W^{(2)}(a^2)}^2W^{(3)}(a^2) \nonu \\
&\!\!   \!\!&
+200{W'(a^2)}^2{W^{(3)}(a^2)}^2
+388{W'(a^2)}^2W^{(2)}(a^2)W^{(4)}(a^2) \nonu \\
&\!\!   \!\!&
-24{W'(a^2)}^3W^{(5)}(a^2)]/(725760{W'(a^2)}^8)\,, \\
E_{-1}^{(3)}
&\!\! = \!\!&
[15{W^{(2)}(a^2)}^3-10W'(a^2)W^{(2)}(a^2)W^{(3)}(a^2) \nonu \\
&\!\!   \!\!&
 +{W'(a^2)}^2W^{(4)}(a^2)]/(6048{W'(a^2)}^7)\,, \\
E_{-2}^{(3)}
&\!\! = \!\!&
[15{W^{(2)}(a^2)}^2-4W'(a^2)W^{(3)}(a^2)]/(6048{W'(a^2)}^6)\,, \\
E_{-3}^{(3)}
&\!\! = \!\!&
W^{(2)}(a^2)/(504{W'(a^2)}^5)\,, \\
E_{-4}^{(3)}
&\!\! = \!\!&
1/(1008{W'(a^2)}^4)\,.
\eea
The validity of this result is checked by applying the simplest
function $W(r_0)=r_0+12\lambda_1r_0^2$, ($\lambda_1\geq -1/48$).
The apparent form of $e^{(3)}(a^2)$ certainly has the predicted
form in \cite{BIZ}.
If the physical fine tuning is performed and the universal terms are
extracted according to the rule (\ref{eqn:ExtractRule}),
one can also confirm the validity of my results by comparing it to
the known results derived by the usual method.


%

%
\end{document}